\documentstyle[12pt]{article}

\newcommand {\ua} {\uparrow}

\newcommand{\be}{\begin{equation}}
\newcommand{\ee}{\end{equation}}  
\newcommand{\eins}{\mbox{$\rule{2.5mm}{0.1mm} 
                          {\hspace{-2.7mm}1} 
                          {\hspace{-0.2mm}\rule{0.07mm}{2.7mm}}$}}
\newcommand{\kslash}{\mbox{$\not{\hspace{-0.8mm}K}$}}          
\newcommand{\pslash}{\mbox{$\not{\hspace{-0.8mm}P}$}}          
\newcommand{\vslash}{\mbox{$\not{\hspace{-0.8mm}v}$}}          
 
\begin{document}
\pagenumbering{arabic}         
\begin{titlepage}
\begin{flushright}
 UTPT-98-08
\end{flushright}
\begin{center} 

  {\large \bf Single Pion Transitions of Charmed Baryons
  
   } 

 \vspace*{0.8cm}

 {\bf Salam Tawfiq \footnote{ A talk given at the MRST `98 conference, 
 `` Toward the Theory of Everything ", May 13-15, 1998. Montr\'eal, Canada. 
 To appear in the proceedings.}
  \hspace*{0.02cm}, \hspace*{0.04cm}  Patrick J. O'Donnell} 

\vspace*{0.4cm}

  Department of Physics, University of Toronto \\
  60 St. George Street, Toronto Ontario, M5S 1A7, Canada

  \vspace*{0.4cm}
  and \\
  \vspace*{0.4cm}

{\large J.G. K\"{o}rner}\\

  \vspace*{0.4cm}

  Institut f\"{u}r Physik, Johannes Gutenberg-Universit\"{a}t\\
  Staudinger Weg 7, D-55099 Mainz, Germany\\
 \vspace*{0.5cm}
{\large May 1998}
       
 \vspace*{1cm}

\end{center} 
\begin {abstract}
 The $SU(2N_{f}) \otimes O(3)$ constituent quark model symmetry of the 
light diquark system are used to analyze single pion transitions of
S-wave to S-wave and P-wave to S-wave heavy baryons. 
We show that the Heavy Quark Symmetry (HQS) coupling factors are given in 
terms of the three independent couplings $g_{\Sigma_Q \Lambda_Q \pi}$, 
 $f_{\Lambda_{Q1} \Sigma_Q \pi}$ and $f_{\Lambda^{*}_{Q1} \Sigma_Q \pi}$.
 Light-Front quark model spin wave functions are, then, employed to calculate 
  these couplings and to predict decay rates of single pion transitions 
  between charm baryon states.
\end{abstract}
\end{titlepage}                             
Heavy Quark Symmetry (HQS) and $SU(2N_f)\times O(3)$ light
diquark symmetry can be used to 
construct heavy baryon spin wave functions a la Bethe-Salpater 
\cite{st:htk,st:kkp}.         
These covariant wave functions were employed \cite{st:hklt} to analyze 
current-induced bottom baryon to charm baryon transitions. Similar
procedure will be followed here to study heavy baryon S-wave to S-wave 
and P-wave to S-wave single pion transitions. In 
single--pion transitions between heavy baryons, the pion is emitted by each of 
the light quarks while the heavy
quark is unaffected. In fact, the heavy baryon velocity will not be 
changed when emitting the pion since it is infinitely massive and will 
not recoil.

In a heavy baryon, a light diquark system with quantum numbers $j^P$ couples 
with a heavy quark with $J_Q^P=1/2^+$ to form a doublet with
 $J^P=(j\pm1/2)$. Heavy quark symmetry allows us to write down a general 
 form for the heavy baryon spin wave functions \cite{st:hks,st:hklt}. Ignoring
 isospin indices, one has 
\begin{equation}\label{st:chigen}
\chi_{\alpha\beta\gamma}=(\phi_{\mu_1\cdots\mu_j})_{\alpha\beta}
\psi_{\gamma}^{\mu_{1} \dots\mu_{j}}(v)\;\;. 
\end{equation}
 Here, $v=\frac{P}{M} $ is the baryon four velocity, $\mu_1\cdots\mu_j$ are
 Lorentz indices, the spinor 
 indices $\alpha$ and $\beta$ refer to the
   light quark system and the index $\gamma$ refers to the heavy quark.
  In the heavy quark limit, the $\chi_{\alpha\beta\gamma}$ satisfy the 
   Bargmann-Wigner equation on the heavy quark index
 \be
 (\vslash)^{\gamma^{\prime}}_{\gamma}\chi_{\alpha\beta\gamma^{\prime}}=
 \chi_{\alpha\beta\gamma} \;\; .
 \ee     
 In general the light degrees of freedom spin wave functions 
$(\phi_{\mu_1\cdots\mu_j})_{\alpha\beta}$
are written in terms of the two bispinors 
$[\chi^0]_{\alpha\beta}$ and 
$[\chi^1_{\mu}]_{\alpha\beta}$. The matrix
$[\chi^0]_{\alpha\beta}=[(\vslash +1)\gamma_5C]_{\alpha\beta}$, projects out     
a spin-0 object, is symmetric when interchanging $\alpha$ and 
 $\beta$. However, $[\chi^1_{\mu}]_{\alpha\beta}=[(\vslash +1)
\gamma_{\perp \mu}C]_{\alpha\beta}$ which projects out a spin-1 object is 
  antisymmetric. Here, $C$ is the charge conjugation operator and 
      $\gamma^{\perp}_{\mu}=\gamma_{\mu}-\vslash v_{\mu}$.
 On the other hand
 the ``superfield" 
 $\psi_{\gamma}^{\mu_{1} \dots\mu_{j}}(v)$ stands for  
  the two spin wave
 functions corresponding to the two heavy quark symmetry degenerate 
 states with spins ($j\pm 1/2$). They are
  generally written in terms of the Dirac spinor $u$ and 
  the Rarita-Schwinger spinor $u_{\mu}$.
  
 The S-wave heavy-baryon spin wave functions are 
 given by 
\be\label{Sstate}
\chi^{\Lambda_Q}_{\alpha\beta\gamma}=(\chi^{0})_{\alpha\beta}u_{\gamma} 
\ee
and
\be  
\chi^{\Sigma_Q}_{\alpha\beta\gamma}=(\chi^{1,\mu})_{\alpha\beta} 
\left\{ \begin{array}{r}\frac{1}{\sqrt{3}}\gamma^{\perp}_{\mu} \gamma_{5}u \\
u_{\mu} \end{array} \right\}_{\gamma} \; .
\ee
 To represent the orbital excitation for P-wave heavy baryon states, one can  
 use the relative momenta
  $K=\frac{1}{\sqrt{6}}(p_1+p_2-2p_3)$ and $k=\frac{1}{\sqrt{2}}(p_1-p_2)$,  
 symmetric and antisymmetric respectively under interchange of the 
 constituent light quark momenta
  $p_1$ and $p_2$. The 
  $\Lambda_{Q1}$ degenerate state spin wave functions can be written as 
 \be \label{Pstate} 
\chi^{\Lambda_{Q1}}_{\alpha\beta\gamma}=({\chi}^{0}K^{\mu})_{\alpha\beta} 
\left\{\begin{array}{r}\frac{1}{\sqrt{3}}\gamma^{\perp}_{\mu} \gamma_{5}u \\
u_{\mu} \end{array}\right\}_{\gamma} \;.
\ee
  S-wave and P-wave heavy baryon spin wave functions 
 are summarized in Table (\ref{st:swf}). 
 \begin{table}
\caption{\label{st:swf}  spin wave functions of S-wave
      and P-wave heavy baryons. The symbol $\{AB\}^0$ refers to a traceless 
      symmetric tensor.}
\vspace{5mm}
\renewcommand{\baselinestretch}{1.2}
\small \normalsize
\begin{center}
\begin{tabular}{|c|c|cc|}
\hline \hline
& $j^{P}$ & $J^{P} $ &  ${\chi}_{\alpha\beta\gamma} $ \\ 
\hline \hline
\multicolumn{4}{|l|} {  S-wave states } \\
 $\Lambda_{Q}$ & $ 0^{+}$ & 
 $\frac{1}{2}^{+}$ & $(\chi^{0})_{\alpha\beta}u_{\gamma}$ \\
\hline
$ \Sigma_{Q} $ & $1^{+}$& 
   $\begin{array}{c}\frac{1}{2}^{+}\\ \frac{3}{2}^{+} \end{array} $ &
         $(\chi^{1,\mu })_{\alpha\beta} 
      \left\{ \begin{array}{r}
             \frac{1}{\sqrt{3}}\gamma^{\perp}_{\mu} \gamma_{5}u\\
             u_{\mu}
         \end{array} \right\}_{\gamma} $ \\ 
\hline\hline
\multicolumn{4}{|l|}{ Symmetric P-wave states}  \\
$ \Lambda_{QK1} $ & $1^{-}$ & 
$\begin{array}{c} \frac{1}{2}^{-} \\\frac{3}{2}^{-} \end{array}$ & 
     $ (\chi^{0} K_{\perp}^{\mu})_{\alpha\beta} \left\{
      \begin{array}{r}
      \frac{1}{\sqrt{3}} \gamma^{\perp}_{\mu} \gamma_{5} u\\
      u_{\mu}
     \end{array} \right\}_{\gamma}$ \\ 
\hline
$\Sigma_{QK0} $ & $0^{-}$ & 
$\frac{1}{2}^{-} $ & $\frac{1}{\sqrt{3}} (\chi^{1,\mu}
 K_{\perp \mu} )_{\alpha\beta} u_{\gamma} $\\ 
\hline
$ \Sigma_{QK1} $ & $1^{-}$ & 
$\begin{array}{c} \frac{1}{2}^{-}  \\ \frac{3}{2}^{-}  \end{array} $ & $
 \frac{i}{\sqrt 2} ( \varepsilon_{\mu\nu\rho\delta}  \chi^{1,\nu} 
  K_{\perp}^{\rho}v^{\delta} )_{\alpha\beta} \left\{ \begin{array}{r}
    \frac{1}{\sqrt{3}} \gamma^{\perp}_{\mu} \gamma_{5} u\\
     u_{\mu} \end{array} \right\}_{\gamma}$ \\ 
\hline
$ \Sigma_{QK2} $& $2^{-}$& 
  $\begin{array}{c} \frac{3}{2}^{-} \\ \frac{5}{2}^{-}  \end{array}$ &
   $ \frac{1}{2}  (\{  \chi^{1,\mu_1} K_{\perp}^{\mu_{2}} \}^{0})_{ \alpha\beta} 
   \left\{ \begin{array}{r}
           \frac{1}{\sqrt{10}}\gamma_{5} \{ \gamma^{\perp}_{ \mu_1 }
              u_{\mu_{2}} \}^0 \\
           u_{\mu_1 \mu_2 }   \end{array} \right\}_{\gamma}$ \\ 
\hline \hline
\multicolumn{4}{|l|}{ Antisymmetric P-wave states } \\
$ \Sigma_{Qk1} $ & $1^{-}$ & 
$\begin{array}{c} \frac{1}{2}^{-} \\\frac{3}{2}^{-} \end{array}$ & 
     $ (\chi^{0} k_{\perp}^{\mu})_{\alpha\beta} \left\{
      \begin{array}{r}
      \frac{1}{\sqrt{3}} \gamma^{\perp}_{\mu } \gamma_{5} u\\
      u_{\mu}
     \end{array} \right\}_{\gamma}$ \\ 
\hline
$\Lambda_{Qk0} $ & $0^{-}$ & 
$\frac{1}{2}^{-} $ & $\frac{1}{\sqrt{3}} (\chi^{1,\mu}  
 k_{\perp\mu} )_{\alpha\beta} 
  u_{\gamma} $ \\ 
\hline
$ \Lambda_{Qk1} $ & $1^{-}$ & 
$\begin{array}{c} \frac{1}{2}^{-} \\ \frac{3}{2}^{-}  \end{array} $ & $ 
 \frac{i}{\sqrt 2} ( \varepsilon_{\mu\nu\rho\delta} \chi^{1,\nu} 
  k_{\perp}^{\rho}v^{\delta} )_{\alpha\beta} \left\{ \begin{array}{r}
    \frac{1}{\sqrt{3}} \gamma^{\perp}_{\mu} \gamma_{5} u\\
     u_{\mu} \end{array} \right\}_{\gamma}$ \\ 
\hline
$ \Lambda_{Qk2} $& $2^{-}$& 
  $\begin{array}{c} \frac{3}{2}^{-} \\ \frac{5}{2}^{-}  \end{array}$ &
   $ \frac{1}{2} (\{ \chi^{1,\mu_1} k_{\perp}^{\mu_{2}} \}^{0})_{ \alpha\beta} 
   \left\{ \begin{array}{r}
           \frac{1}{\sqrt{10}}\gamma_{5} \{ \gamma^{\perp}_{ \mu_1 }
              u_{\mu_{2}} \}^0 \\
           u_{\mu_1 \mu_{2}} \end{array} \right\}_{\gamma}$ \\ 
\hline \hline
\end{tabular}
\renewcommand{\baselinestretch}{1}
\small \normalsize
\end{center}
\end{table}
The heavy baryon total wave functions
are constructed ensuring overall symmetry with respect to $flavour\otimes
spin\otimes orbital$.

The one--pion transition amplitudes between 
heavy baryons can then be written as 
\begin{eqnarray}
M^{\pi}&=& \langle \pi (\vec{p}), B_{Q}(v) \mid T \mid B_{Q}(v) \rangle
\nonumber\\[2mm]
&=&{\bar \psi}_{2}^{\nu_{1} \dots \nu_{j_{2}}}(v) \psi_{1}^{\mu_{1} \dots
\mu_{j_{1}}}(v) {\cal M}_{\mu_{1} \dots
\mu_{j_{1}};\nu_{1} \dots \nu_{j_{2}}} \,.\label{st:trans}
\end{eqnarray}
The light diquark tensors
${\cal M}_{\mu_1 \dots \mu_{j_1};\nu_1 \dots \nu_{j_2}}$
of rank $(j_{1}+j_{2})$, describe $j_1\rightarrow j_2+\pi$ transitions, should 
have the correct parity and project out the appropriate partial wave amplitude.
    
HQS predicts that S-wave to S-wave transitions involve 
two $p-$wave coupling constants. However, each of the single pion
transitions from the K-multiplet and from the k-multiplet down to the ground
state are determined in terms
of seven coupling constants. In fact, there are three $s-$wave and 
four $d-$wave couplings for each. Matrix elements of these transitions 
are explicitly given in \cite{st:hks}. 

 To go beyond HQS predictions, we invoke the $SU(6)\otimes O(3)$ symmetry
of the light degrees of freedom to calculate the light-side transition matrix 
elements ${\cal M}$. In S-wave to S-wave single-pion decays, the transitions 
involved are $1^+\rightarrow 0^+$ and $1^+\rightarrow 1^+$ with 
light diquark transition tensors given by
\be\label{st:stos}
 {\cal M}_{\mu_1;\nu_{j_2}}=
\left( {\bar \phi_{\nu_{j_2}}} \right)^{\alpha\beta}
\left({\cal O}\right)^{\alpha^{'}\beta^{'}}_{\alpha\beta}
\left({\phi}_{\mu_1}\right)_{\alpha^{'}\beta^{'}},  
\ee  
where the operator ${\cal O}$ is given in terms of an overlap integral which 
is unknown. Constraints on the operator $\cal O$ come from parity 
conservation and from the partial wave involved in the emission 
process. Since $l_{\pi}=1$, therefore, 
 it is easy to show that $\cal O$ must be a pseudoscalar 
 operator involving one power of the pion momentum $p$. In the
constituent quark model the pion is emitted by one of the light 
quarks, hence, the transition operator ${\cal O}$  
must be a one-body operator. Possible two-body emission operators are
non leading in large-$N_C$ \cite{st:witt-oth} and are thus neglected in the 
constituent quark model approach \cite{st:cgkm}.
One then has the unique operator 
\be\label{st:ostos}
\left({\cal O}(p)\right)^{\alpha\beta}_{\alpha^{'}\beta^{'}} =\frac{1}{2}
\left((\gamma_\sigma\gamma_{5})^{\alpha}_{\alpha^{'}}
 \otimes (\eins)^{\beta}_{\beta^{'}}
          + \,(\eins )^{\alpha}_{\alpha^{'}}\otimes (\gamma_\sigma\gamma_5)
^{\beta}_{\beta^{'}}\right)
 f_{p}\,p_{ \perp}^{\sigma}\,
\ee  
The relevant transition tensors for P-wave to S-wave single-pion decays, 
which involve $1^-\rightarrow \{0^+,1^+\}$, $0^-\rightarrow 0^+$ and 
$2^-\rightarrow \{0^+,1^+\}$, are given by
\be\label{st:ptos}
 {\cal M}_{\mu_1\cdots\mu_{j_1};\nu_{j_2}}=\sum_{l_{\pi}=0,2}
\left(\bar{\phi}_{\nu_{\j_2}}\right)^{\alpha\beta}
\left({\cal O}_{\lambda}^{(l_{\pi})}\right)
^{\alpha^{'}\beta^{'}}_
{\alpha\beta}\left({\phi}^{\lambda}_{\mu_1\cdots\mu_{j_1}}
\right)_{\alpha^{'}\beta^{'}}\,,  
\ee                                                                          
 the appropriate operators for these transitions are given by 
\be
\left({\cal O}_{\lambda}(p)\right)^{\alpha\beta}_{\alpha^{'}\beta^{'}} =
\frac{1}{2}\left((\gamma^\sigma\gamma_5)^{\alpha}_{\alpha^{'}}
 \otimes (\eins)^{\beta}_{\beta^{'}}
          \pm \,(\eins )^{\alpha}_{\alpha^{'}}\otimes (\gamma^\sigma\gamma_5)
^{\beta}_{\beta^{'}}\right)
\left( f_s\;g_{\sigma\lambda}+ f_d\;P_{\sigma\lambda}\right),                  
\label{st:Ktrans}
\ee
 with, $P_{\sigma\lambda}(p)=p_{\perp \sigma}p_{\perp \lambda}-
\frac{1}{3}p_{\perp}^2 g_{\perp \sigma \lambda}$. The plus sign has to be 
used for transitions from the Symmetric (K-multiplet) and the minus one for 
transitions from the Antisymmetric (k-multiplet). P-wave to P-wave 
transitions were analyzed in \cite{st:hks} and the generalization 
to transitions involving higher orbital excitations is straightforward.

The matrix elements, Eq. (\ref{st:stos}) and Eq. (\ref{st:ptos}), of the
operators Eq. (\ref{st:ostos}) and Eq. (\ref{st:Ktrans}), can
be readily evaluated using the light diquark
spin wave functions in Table(\ref{st:swf}). The two couplings of the ground 
state transitions are not independent. They are, actually, related to the 
single p-wave coupling $f_{p}$ by
\be\label{st:fpi}
f_p^1=-f_p^2=f_p \, .
\ee
Using PCAC the coupling constant $f_p$ can be related to the
axial vector current coupling strength $g_A$, one obtains
$f_{p}=g_A/f_{\pi}$. 
 
For P-wave (K-multiplet) to S-wave transitions, the evaluation of the matrix
elements leads to the following relations 
\begin{eqnarray}\label{st:fsK}
f_s^{1(K)}=f_s\;\; ; \;\; f_s^{2(K)}=-\sqrt{3}f_s
\;\; ; \;\;f_s^{3(K)}=\sqrt{2}f_s \\
f_d^{1(K)}=f_d\;\; ; \;\; f_d^{3(K)}=-\frac{1}{\sqrt{2}}f_d
\;\; ; \;\;  f_d^{4(K)}=-f_d\;\; ; \;\;  f_d^{5(K)}=f_d\label{st:fdK}
\end{eqnarray}   
The number of independent coupling constants, therefore, has been reduced 
from seven to the two constituent quark model $s$-wave and $d$-wave coupling
factors $f_{s}$ and $f_{d}$. Similar relations, with two different 
couplings, hold for transitions from the P-wave (k-multiplet) to S-wave.

The first important conclusion we have reached so far is that, S-wave 
to S-wave and P-wave (K-multiplet) 
to S-wave single pion transitions are given in terms of the three independent 
couplings $f_p$, $f_s$ and $f_d$. For charmed baryons, they can be 
identified by the three strong couplings $g_{\Sigma_c \Lambda_c \pi}$, 
 $f_{\Lambda_{c1} \Sigma_c \pi}$ and $f_{\Lambda^{*}_{c1} \Sigma_c \pi}$ 
respectively. We would like to mention that, after taking into account 
the different normalizations,   
the results Eqs. (\ref{st:fpi}) and (\ref{st:fsK}--\ref{st:fdK}) are in 
agreement with corresponding results using HHCPT \cite{st:yp}.

  The three independent couplings can be written in terms of Light-Front (LF) 
  \cite{st:lc} matrix elements 
 of the strong transition current ${\hat j}_{\pi}(q)$ between LF heavy 
 baryon helicity states.
 Working in the Drell-Yan frame, we get \cite{st:tok}
\begin{equation}\label{st:g2}
 g_{\Sigma_c \Lambda_c \pi}=
 -\frac{2\sqrt{3M_{\Lambda_c}M_{\Sigma_c}}}{(M_{\Sigma_c}^2-M_{\Lambda_c}^2)}
 \langle \Lambda (P^{\prime},{\ua})
|{\hat j}_{\pi}(0)|\Sigma(P,{\ua}) \rangle
\end{equation}
 \begin{equation}\label{st:h2}
 f_{\Lambda_{c1} \Sigma_c \pi}=
 \langle\Sigma (P^{\prime},{\ua})
 | {\hat j}_{\pi}(0)|\Lambda_{c1}(P,{\ua}) \rangle
 \; ,
 \end{equation}
 and
 \be\label{st:h8}
 f_{\Lambda^{*}_{c1} \Sigma_c \pi}=
 \frac{3\sqrt{2}}{ (M_{\Lambda^{*}_{c1}}-M_{\Sigma})^2}
 \frac{M_{\Lambda^{*}_{c1}}^2}{ (M_{\Lambda^{*}_{c1}}^2-M_{\Sigma_c}^2)} 
\langle \Sigma (P^{\prime},{\ua})
 |{\hat j}_{\pi}(0)|\Lambda^{* }_{c1}(P,\frac{1}{2}) \rangle
 \ee
  In the LF formalism the total baryon spin-momentum distribution function 
  can be written in the following general form  
\be\label{st:psitot}
\Psi(x_i,{\bf p}_{\perp i},\lambda_i;{\lambda})=
\chi(x_i,{\bf p}_{\perp i},\lambda_i;{\lambda}) \psi(x_i,{\bf p}_{\perp i}).
 \ee
 Here, $\chi(x_i,{\bf p}_{\perp i},\lambda_i;{\lambda})$ and 
 $\psi(x_i,{\bf p}_{\perp i})$ represent the spin and momentum distribution 
 functions respectively. 
 Assuming factorization of longitudinal and
 transverse momentum distribution functions, one can write 
 \be\label{st:mom} 
 \psi(x_i,{\bf p}_{\perp i})=\prod_{i=1}^{3} \delta(x_i-\bar{x}_i)
 exp\left[-\frac{ \stackrel{\rightarrow}{ 
{\bf k}} ^2}{2\alpha_{\rho}^2}
  -\frac{ \stackrel{\rightarrow}{{\bf K}}^2}{2\alpha^2_{\lambda}}
  \right]\; .
\ee
 The longitudinal momentum distribution functions are approximated by 
 Dirac-delta functions which are peaked at the constituent quark 
 longitudinal momenta mean values $\bar{x}_i=\frac{m_i}{M}$. This assumption 
 is justified since in the weak binding \cite{st:hkt} and the valence 
 \cite{st:vqm} approximations, the constituent quarks are moving 
 with the same velocity inside the baryon.  The heavy baryons spin wave 
 functions, which are the LF generalization of the conventional constituent 
 quark model spin-isospin functions, are explicitly given by \cite{st:tok}  
 \be\label{chilam}
 \chi^{\Lambda_Q}(x_i,{\bf p}_{\perp i},\lambda_i;\lambda)=\bar{u}
( p_{1},{\lambda_1})[(\pslash +M_{\Lambda})\gamma_5]\nu
( p_{2},{\lambda_2})
 \bar{u}(p_3,{\lambda_3})u(P,{\lambda}).
  \ee
 For the $\Sigma_Q$-like baryons, one has
 \be\label{chisig2}
 \chi^{\Sigma_Q}(x_i,{\bf p}_{\perp i},\lambda_i;{\lambda})=\bar{u}
 (p_1,{\lambda_1})[(\pslash +M_{\Lambda})\gamma_{\perp}^{\mu}]\nu
 (p_2,{\lambda_2})
 \bar{u}(p_3,{\lambda_3})\gamma_{\perp \mu}\gamma_5 
 u(P,{\lambda}),
 \ee
  The excited states $\Lambda_{Q1}$, with $J^P=\frac{1}{2}^-$, and 
 $\Lambda^{*}_{Q1}$, with $J^P=\frac{3}{2}^-$,  
  have spin functions of the forms 
\be\label{chilam12}
 \chi^{\Lambda_{QK1}}(x_i,{\bf p}_{\perp i},\lambda_i;{\lambda})=
 \bar{u}
(p_1,{\lambda_1})[(\pslash +M_{\Lambda_{c1}})\gamma_5]\nu(p_2,{\lambda_2})
 \bar{u}(p_3,{\lambda_3})\kslash  \gamma_5 
 u(P,{\lambda}),
\ee
and 
\be\label{chilam32}
\chi^{\Lambda^{*}_{QK1}}(x_i,{\bf p}_{\perp i},\lambda_i;{\lambda})=
 \bar{u}
 (p_1,{\lambda_1})[(\pslash +M_{\Lambda^{*}_{c1}})\gamma_5]\nu
 (p_2,{\lambda_2})
 \bar{u}(p_3,{\lambda_3}) K_{\mu}  
 u^{\mu}(P,\lambda) \;\; .
\ee
  The three charmed baryons strong couplings 
 $g_{\Sigma_c\Lambda_c\pi}$, $f_{\Lambda_{c1}\Sigma_c\pi}$ and 
 $f_{\Lambda^{*}_{c1}\Sigma_c\pi}$ are calculated 
 \footnote{ The numerical values for 
 the constituent quark masses and the oscillator couplings 
 are taken to be $m_u=m_d=0.33 {\rm \; GeV}$, 
 $m_c=1.51 {\rm \; GeV}$,
  $\alpha_{\rho}=0.40\; {\rm GeV}/c$ and $\alpha_{\lambda}=0.52 
 \; {\rm GeV}/c$. The charmed baryon masses will be
  taken from Table 1 of \cite{st:cf}.} 
 to be 
 \be\label{st:theory}
 g_{\Sigma_c\Lambda_c\pi}=6.81\; {\rm GeV}^{-1} \; \; , \;\; 
   f_{\Lambda_{c1}\Sigma_c\pi}=1.16 \;\; ,\;\;
 f_{\Lambda^{*}_{c1}\Sigma_c\pi}=0.96\times 10^{-4}\; {\rm MeV}^{-2} \;. 
\ee
 These values can be used to determine the corresponding HHCPT couplings, one 
 gets  
\be
 g_2=0.52  \; \; {\rm ,} \;\;
 h_2=0.54 \;\; {\rm ,} \;\; h_8=3.33\times 10^{-3}{\rm MeV}^{-1}\; .
 \ee 
 Assuming that the 
  width of 
 $\Sigma_c$, $ \Lambda_{c1} $ and 
 $ \Lambda^{*}_{c1} $ are saturated by strong decay channels one can 
 estimate the values 
 of the three couplings using the experimental decay rates. 
  CLEO \cite{st:exp} results for 
  $\Gamma_{ \Sigma_c^{* ++}\rightarrow \Lambda_c^{+} \pi^{+}}=
  17.9^{+3.8}_{-3.2} \; {\rm MeV} $ and
  $\Gamma_{ \Sigma_c^{*0}\rightarrow \Lambda_c^{+} \pi^{-}}=
  13.0^{+3.7}_{-3.0}\; {\rm MeV}$
 can be used to determine 
  the coupling $g_{\Sigma_c\Lambda_c\pi}$. One, therefore, respectively gets 
\be\label{st:g-value}
  g_{\Sigma_c\Lambda_c\pi}=8.03^{+1.97}_{-1.92} \; {\rm GeV}^{-1} \;
\ee
 and
\be
g_{\Sigma_c\Lambda_c\pi}=6.97^{+1.84}_{-1.74}\; {\rm GeV}^{-1} \;
\ee
  To estimate $f_{\Lambda_{c1}\Sigma\pi}$ we use the
    Particle Data Group \cite{st:PDG} average
   value for 
   $ \Lambda_{c_1}(2593)$ width 
   ($\Gamma_{\Lambda_{c_1}(2593)}=3.6^{+2.0}_{-1.3}\; {\rm MeV} $) to obtain  
 \be 
  f_{\Lambda_{c1}\Sigma\pi}=1.11^{+0.31}_{-0.20}.
 \ee
  Finally, taking the upper bound on the $\Lambda^{+}_{c_1}(2625)$ width 
  obtained by CLEO \cite{st:exp} ($\Gamma_{\Lambda^{+}_{c_1}(2625)}
  < 1.9\; {\rm MeV}$) one gets
  \be \label{st:f-value}
  f_{\Lambda^{*}_{c1}\Sigma\pi}=1.66\times 10^{-4}\; {\rm MeV}^{-2} \;.
  \ee
  The LF quark model predictions for the numerical 
 values of the single-pion couplings Eq. (\ref{st:theory}) are in good 
 agreement with estimates obtained using the 
 available experimental data Eqs. (\ref{st:g-value}-\ref{st:f-value}).
 
  We are now in a position to 
 predict charmed baryons strong decay rates using the general formula   
  \be\label{st:rate}
  \Gamma = \frac{1}{2J_{1}+1} \quad \frac{ \mid \vec{q} \mid}{8 \pi
  M_{B_Q}^{2}}\sum_{spins} \mid M^{\pi} \mid^{2},
  \ee
  with $\mid \vec{q} \mid$ being the pion momentum in the rest frame 
  of the decaying baryon.
 The numerical values for S-wave to S-wave and P-wave (K-multiplet) to S-wave 
 single pion decay rates and the updated experimental values 
   of the Review of Particle Physics \cite{st:PDG} are summarized in 
   Table \ref{st:rates}. 
\begin{table}
\caption{\label{st:rates} Decay rates for charmed baryon states.}
 \vspace{5mm}
 \renewcommand{\baselinestretch}{1.2}
 \small \normalsize
 \begin{center}
 \begin{tabular}{|c|c|c|}
 \hline \hline
 $B_Q\rightarrow B^{\prime}_{Q}\pi$ & $\Gamma \; ({\rm MeV}) $ & 
 $ \Gamma_{expt.} \; ({\rm MeV}) $  \\
 \hline\hline 
\multicolumn{3}{|l|}{Ground state transitions} \\ 
\hline  
$\Sigma^{+}_{c} \rightarrow \Lambda_c\pi^{0} $ &  $ 1.70  $ & $  $   \\ 
$\Sigma^{0}_{c} \rightarrow \Lambda_c\pi^{-} $ &  $ 1.57   $ & $  $   \\ 
$\Sigma^{++}_{c} \rightarrow \Lambda_c\pi^{+} $ &  $ 1.64 $ & $  $   \\ 
\hline 
 $\Sigma^{*0}_{c} \rightarrow \Lambda_c\pi^{-} $ &  $ 12.40 $ &  
 $ 13.0^{+3.7}_{-3.0} $   \\     
 $\Sigma^{* ++}_{c} \rightarrow \Lambda_c\pi^{+} $ &  $ 12.84 $ &  
 $ 17.9^{+3.8}_{-3.2} $   \\  
\hline
$\Xi^{*0}_{c} \rightarrow \Xi^{0}_c\pi^{0} $ &  $  0.72 $ & $ <5.5 $   \\
$\Xi^{*0}_{c} \rightarrow \Xi^{+}_c\pi^{-} $ &  $  1.16 $ &   \\
\hline
$\Xi^{*+}_{c} \rightarrow \Xi^{0}_c\pi^{+} $ &  $ 1.12  $ & $ <3.1 $    \\
$\Xi^{*+}_{c} \rightarrow \Xi^{+}_c\pi^{0} $ &  $ 0.69 $ & $  $   \\
\hline \hline 
\multicolumn{3}{|l|}{P-wave to S-wave transitions} \\ 
\hline  
$\Lambda_{c1}(2593) \rightarrow \Sigma^{0}_c\pi^{+} $ & $2.61  $ & $ $  \\
$\Lambda_{c1}(2593) \rightarrow \Sigma^{+}_c\pi^{0} $ & $1.73  $ & 
$ 3.6^{+2.0}_{-1.3}$  \\
$\Lambda_{c1}(2593) \rightarrow \Sigma^{++}_c\pi^{-} $ & $2.15  $ & $ $  \\
\hline  
$\Lambda^{*}_{c1}(2625) \rightarrow \Sigma^{0}_c\pi^{+} $ & $ 0.77 $ & $  $ \\
$\Lambda^{*}_{c1}(2625) \rightarrow \Sigma^{+}_c\pi^{0} $ & $ 0.69 $ & 
$\Gamma_{\Lambda^{*}_{c1}}<1.9 $  \\
$\Lambda^{*}_{c1}(2625) \rightarrow \Sigma^{++}_c\pi^{-} $ & $ 0.73 $ & $ $  \\
 \hline 
$\Xi^{*}_{c1}(2815) \rightarrow \Xi^{*0}_c\pi^{+} $ &  $ 4.84  $ &  
$ \Gamma_{\Xi^{*}_{c1}}<2.4 $   \\
$\Xi^{*}_{c1}(2815) \rightarrow \Xi^{*+}_c\pi^{0} $ &  $ 2.38  $ & $  $   \\
\hline 
$\Xi^{*}_{c1}(2815) \rightarrow \Xi^{0}_c\pi^{+} $ & $ 0.30 $ & $ $  \\
$\Xi^{*}_{c1}(2815) \rightarrow \Xi^{+}_c\pi^{0} $ & $ 0.15 $ & $ $  \\
\hline 
 $\Sigma_{c0}(2760) \rightarrow \Lambda_c\pi $ &  $ 110.36   $ & $  $ \\ 
 \hline 
 $\Sigma_{c1}(2770) \rightarrow \Sigma_c\pi $ &  $ 50.92   $ & $  $  \\
\hline 
$\Sigma_{c2}(2800) \rightarrow \Sigma^{*}_c\pi $ &  $ 50.21 $ & $  $ \\
\hline \hline  
\end{tabular}
 \renewcommand{\baselinestretch}{1}
 \small \normalsize
 \end{center}
 \end{table}              
 To predict the total decay width of these states, one has to include 
 decay rates for the two-pion transitions reported by \cite{st:yp}.
 Table \ref{st:rates} shows that most of the predicted decay 
 widths agree quite well or they are within the range of the corresponding
 experimental data. We, also, notice that the $\Sigma_{c0}(2760)$ ,
 $\Sigma_{c1}(2770)$ and $\Sigma_{c2}(2800)$ widths are relatively broad and 
 it might be difficult to measure them experimentally.  

 To summarize, we have used
 the $SU(2N_f)\times O(3)$ symmetry, of the light 
 diquark system, to reduce the number of HQS 
 coupling factors of heavy baryon single-pion decays. These result, which are 
 obtained using covariant spin wave functions for the light diquark 
 system, agree with the HHCPT \cite{st:yp}. 
 We also calculated the 
 three independent couplings $g_{\Sigma_c \Lambda_c \pi}$, 
 $f_{\Lambda_{c1} \Sigma_c \pi}$ and $f_{\Lambda^{*}_{c1} \Sigma_c \pi}$
 using a Light-Front (LF) quark model functions.               
  Most of the predicted decay rates agree with the available experimental data. 
 Like other models, our numerical result will depend on the values of the 
 the constituent quark masses and the 
 harmonic oscillator constants $\alpha_{\rho}$ and $\alpha_{\lambda}$ which are 
 free parameters.   
 \section*{Acknowledgments}
   One of us S. T. would like to thank 
  Patrick J. O'Donnell and the Department of Physics, University of 
  Toronto for hospitality. This 
  research was supported in part by the National Sciences and Engineering 
  Research Council of Canada.


\begin{thebibliography}{999}
\bibitem{st:htk} F. Hussain, G. Thompson J.G. K\"orner,  preprint IC/93/314,
MZ-TH/93-23 and hep-ph/9311309, to appear in the proceedings of the 6th.
Regional Conference in Mathematical Physics, Islamabad, February 1994.   
\bibitem{st:kkp} J.G. K\"orner, M. Kr\"amer and D. Pirjol, Progr. Part. Nucl.
Phys., Vol. 33 (1994)787.  
\bibitem{st:hklt} F. Hussain, J.G. K\"orner, J. Landgraf and Salam Tawfiq,
Z. Phys. {\bf C69} (1996) 655.
\bibitem{st:hks} F. Hussain, J.G. K\"orner and Salam Tawfiq,
ICTP preprent, IC/96/35 and Mainz preprint MZ-TH/96-10, 1996.
\bibitem{st:cho} P. Cho, Nucl Phys. {\bf B396}(1993)183; Phys. Rev. {\bf D50} (1994)3295.
\bibitem{st:hdh} M-Q Huang, Y-B Dai and C-S Huang, Phys. Rev. {\bf D52}
(1995) 3986.
\bibitem{st:yp} D. Pirjol and T. M. Yan, Phys. Rev. {\bf D56}(1997)5483. 
\bibitem{st:cf} G. Chiladze and A. Falk, Phys. Rev. {\bf D56}(1997)6738.
\bibitem{st:witt-oth} E. Witten, Nucl.Phys. {\bf B223}(1983)483;
C. Carone, H. Georgi and S. Osofski, Phys. Lett. {\bf 322B}(1994)483; 
M. Luty and
J. March-Russel, Nucl. Phys. {\bf B246}(1994)71; R. F. Dashen, E. Jenkins and
A. V. Manohar, Phys. Rev. {\bf D49}(1994)4713; R. F. Dashen, E. Jenkins and
A.V.Manohar, Phys. Rev. {\bf D51}(1995)3697.
\bibitem{st:cgkm} C. Carone, H. Georgi, L. Kaplan and D. Morin, Phys. Rev.
{\bf D50}(1994)5793.      
\bibitem{st:ycclly}T. M. Yan et. al., Phys. Rev. {\bf D}46 (1992)1148;
Erratum, to be published.
\bibitem{st:lc} For a recent review and references see S. J. Brodsky, H-S Pauli
  and S. S. Pinsky, hep-ph/9705477, 1997c
\bibitem{st:tok} Salam Tawfiq, Patrick J. O'Donnell and J.G. K\"orner, 
hep-ph/9803246, University of Toronto preprint UTPT-98-03 and 
Mainz preprint MZ-TH/98-08, 1998, to appear in Phys. Rev. {\bf D.}.
\bibitem{st:hkt} F. Hussain, J.G. K\"orner and G. Thompson, Ann. Phys. 
 {\bf C59}(1993)334. 
\bibitem{st:vqm} Z. Dziembowski, Phys. Rev {\bf D37}(1988)778;
  H. J. Weber, Phys. Lett {\bf B209}(1988)425; 
  Z. Dziembowski and H. J. Weber,  Phys. Rev {\bf D37}(1988)1289;  
   W. Konen and H. J. Weber, Phys. Rev. {\bf D41}(1990)2201.
\bibitem{st:exp} G. Brandenburg et.al., CLEO Coll., 
 Phys. Rev. Lett {\bf 78}(1997)2304; L. Gibbons et.al., CLEO Coll.,
 Phys. Rev. Lett {\bf 77}(1996)810; 
 P. L. Frabetti et.al, E687 Coll., Phys. Lett {\bf B365}(1996)461;
 Phys. Rev. Lett. {\bf 72} (1994) 961;
 K. W. Edwards et.al., CLEO Coll., Phys. Rev. Lett. {\bf 74} (1995) 3331;
 P. Avery, el.al., CLEO Coll., Phys. Rev. Lett.
 {\bf 75} (1995) 4364;
 H. Albrecht et.al., ARGUS Coll., Phys. Lett. {\bf B317} (1993) 227.
\bibitem{st:PDG} R. M. Barnett et.al., Phys. Rev. {\bf D54} (1996)1 and 1997 
 off-year partial updated for the 1998 edition, http://pdg.lbl.gov/.
\end{thebibliography}
\end{document}